\input{aipcheck.tex}
\documentclass[article]{aipproc}
\layoutstyle{6x9}
\usepackage{feynarts}
\newcommand\doingARLO[2][]{%
  \ifx\mmref\undefined #1\else #2\fi
}

\begin{document}

\title{Mixing and lifetimes of $b$-hadrons}


\author{Alexander J. Lenz}{
  address={Fakult{\"a}t f{\"u}r Physik, Universit{\"a}t Regensburg, 
           93040 Regensburg, Germany},
  email={alexander.lenz@physik.uni-regensburg.de}}

\begin{abstract}
We review the status of mixing and lifetimes of $b$-hadrons.
We will show that  $\Delta \Gamma / \Delta M$, $a_{sl}$ and $\phi$
are better suited to search for new physics effects than $\Delta M$ alone,
because of our poor knowledge of the decay constants.
The theoretical precision in the determination of $\Gamma_{12} / M_{12}$ 
- which contains all information on $\Delta \Gamma / \Delta M$, $a_{sl}$ 
and $\phi$ - can be tested directly by investigating the lifetimes of 
$b$-hadrons, because both quantities rely on the same theoretical footing.
In particular we will also present a numerical estimate for the lifetime
of the $\Xi_b$-baryon.
\end{abstract}

\date{\today}

\maketitle

\section{Introduction - Theoretical Tools}
In this section we briefly discuss the principles of the calculation of 
physical quantities like lifetimes of $b$-hadrons $\tau$, 
the mass difference in the neutral
B-meson system $\Delta M$, the decay rate difference in the neutral
B-meson system $\Delta \Gamma$, the semi-leptonic CP-asymmetry $a_{sl}$
and the mixing phase $\phi$. 
These quantities are currently measured at the B-factories and at the 
TeVatron (see e.g. \cite{Cern03}) and they will be measured at the
LHC \cite{CERN08} or at a Super-B-factory \cite{SuperB07} with high 
precision.
They are defined as (see e.g. \cite{LN06} for more details): 
\vspace{-0.5cm}
\begin{eqnarray}
\frac{1}{\tau} = \sum \limits_X \Gamma (B \to X) \, , &&
\\
\Delta M      =  M_H - M_L = 2 | M_{12} | \, ,
&&
\Delta \Gamma =  \Gamma_L - \Gamma_H = 2 |\Gamma_{12}| \cos (\phi) \, ,
\\
a_{sl}        = \Im \left( \frac{\Gamma_{12}}{M_{12}}\right) \, ,
&&
\phi          = \arg \left( - \frac{M_{12}}{\Gamma_{12}}\right) \, .
\end{eqnarray} 
For the lifetimes one has to sum the decay rates into all possible 
final states $X$.
The mixing stems from so-called box diagrams (see below). 
$M_{12}$ is the dispersive part (sensitive to heavy internal particles) 
and $\Gamma_{12}$ is the absorptive part (sensitive to light internal 
particles) of these box diagrams.
In the standard model $\tau = 1/\Gamma$, 
$M_{12}$ and $\Gamma_{12}$ are given by the 
following diagrams (as an example we draw the diagrams for the $B_s$-meson):
\vspace{-0.5cm}
\begin{center}
\begin{feynartspicture}(420,130)(3,1)
\FADiagram{}
\FAProp(4.,18.)(10.,18.)(0.,){/Straight}{1}
\FALabel(7.,18.52)[b]{$b$}
\FAProp(10.,18.)(18.,18.)(0.,){/Straight}{1}
\FALabel(16.,18.52)[b]{$c,u$}
\FAProp(10.,18.)(13.,12.)(0.,){/Sine}{0}
\FALabel(9.48,14.48)[tr]{W}
\FAProp(13.,12.)(18.,12.)(0.,){/Straight}{-1}
\FALabel(15.5,13.02)[b]{$\bar{c},\bar{u}$}
\FAProp(13.,12.)(18.,8.5)(0.,){/Straight}{1}
\FALabel(14.8389,9.0685)[tr]{$s,d$}
\FAProp(3.,19.)(3.,7.)(0.,){/Straight}{0}
\FALabel(2.18,11.5)[r]{$\Gamma = \int \sum \limits_X$}
\FAProp(18.5,19.)(18.5,7.)(0.,){/Straight}{0}
\FALabel(19.02,19.5)[l]{2}
\FAVert(10.,18.){0}
\FAVert(13.,12.){0}
\FADiagram{}
\FAProp(2.,15.)(7.,15.)(0.,){/Straight}{1}
\FALabel(4.5,16.02)[b]{$b$}
\FAProp(7.,15.)(13.,15.)(0.,){/Straight}{1}
\FALabel(10.,16.02)[b]{$t$}
\FAProp(13.,15.)(18.,15.)(0.,){/Straight}{1}
\FALabel(15.5,16.02)[b]{$s$}
\FAProp(2.,9.)(7.,9.)(0.,){/Straight}{-1}
\FALabel(4.5,8.18)[t]{$\bar{s}$}
\FAProp(7.,9.)(13.,9.)(0.,){/Straight}{-1}
\FALabel(10.,8.18)[t]{$\bar{t}$}
\FAProp(13.,9.)(18.,9.)(0.,){/Straight}{-1}
\FALabel(15.5,8.18)[t]{$\bar{b}$}
\FAProp(7.,9.)(7.,15.)(0.,){/Sine}{0}
\FALabel(9.58,12.)[r]{W}
\FALabel(3.38,12.)[r]{$, \, \, M_{12}=$}
\FAProp(13.,9.)(13.,15.)(0.,){/Sine}{0}
\FALabel(13.82,12.)[l]{W}
\FAVert(7.,15.){0}
\FAVert(13.,15.){0}
\FAVert(7.,9.){0}
\FAVert(13.,9.){0}
\FADiagram{}
\FAProp(2.,15.)(7.,15.)(0.,){/Straight}{1}
\FALabel(4.5,16.02)[b]{$b$}
\FAProp(7.,15.)(13.,15.)(0.,){/Straight}{1}
\FALabel(10.,16.02)[b]{$c,u$}
\FAProp(13.,15.)(18.,15.)(0.,){/Straight}{1}
\FALabel(15.5,16.02)[b]{$s$}
\FAProp(2.,9.)(7.,9.)(0.,){/Straight}{-1}
\FALabel(4.5,8.18)[t]{$\bar{s}$}
\FAProp(7.,9.)(13.,9.)(0.,){/Straight}{-1}
\FALabel(10.,8.18)[t]{$\bar{c}, \bar{u}$}
\FAProp(13.,9.)(18.,9.)(0.,){/Straight}{-1}
\FALabel(15.5,8.18)[t]{$\bar{b}$}
\FAProp(7.,9.)(7.,15.)(0.,){/Sine}{0}
\FALabel(9.58,12.)[r]{W}
\FALabel(2.98,12.)[r]{$, \, \, \Gamma_{12}=$}
\FAProp(13.,9.)(13.,15.)(0.,){/Sine}{0}
\FALabel(13.82,12.)[l]{W}
\FAVert(7.,15.){0}
\FAVert(13.,15.){0}
\FAVert(7.,9.){0}
\FAVert(13.,9.){0}
\end{feynartspicture}
\end{center}
\vspace{-1.5cm}
All these quantities are triggered by weak decays, in particular
by the exchange of heavy $W$-bosons and the top-quark. Using the fact that 
these particles are much heavier than the b-quark ($m_t, m_W \gg m_b$)
one can integrate them out by performing an operator product expansion
(OPE I), see e.g. \cite{Buras98} for a nice introduction.
In the resulting effective theory the standard model diagrams are rewritten
in a product of perturbative Wilson coefficients and new operators, they
now look like that:
\begin{center}
\begin{feynartspicture}(420,130)(3,1)
\FADiagram{}
\FAProp(4.,18.)(11.5,15.)(0.,){/Straight}{1}
\FALabel(8.1213,16.9484)[b]{$b$}
\FAProp(11.5,15.)(18.,18.)(0.,){/Straight}{1}
\FALabel(16.6268,17.766)[br]{$c,u$}
\FAProp(11.5,15.)(18.,12.)(0.,){/Straight}{-1}
\FALabel(14.8985,14.3819)[bl]{$\bar{c},\bar{u}$}
\FAProp(11.5,15.)(18.,8.5)(0.,){/Straight}{1}
\FALabel(13.8175,10.792)[tr]{$s,d$}
\FAProp(3.,19.)(3.,7.)(0.,){/Straight}{0}
\FALabel(2.18,11.5)[r]{$\Gamma = \int \sum \limits_X$}
\FAProp(18.5,19.)(18.5,7.)(0.,){/Straight}{0}
\FALabel(19.02,19.5)[l]{2}
\FAVert(11.5,15.){2}
\FADiagram{}
\FAProp(2.,15.)(10.,12.)(0.,){/Straight}{1}
\FALabel(6.5266,14.4244)[b]{$b$}
\FAProp(10.,12.)(18.,15.)(0.,){/Straight}{1}
\FALabel(13.4733,14.4244)[b]{$s$}
\FAProp(2.,9.)(10.,12.)(0.,){/Straight}{-1}
\FALabel(6.4564,9.7627)[t]{$\bar{s}$}
\FAProp(10.,12.)(18.,9.)(0.,){/Straight}{-1}
\FALabel(13.5435,9.7627)[t]{$\bar{b}$}
\FALabel(3.5,12.)[r]{$, \, \, \, M_{12} = $}
\FAVert(10.,12.){3}
\FADiagram{}
\FAProp(2.,15.)(7.,12.)(0.,){/Straight}{1}
\FALabel(4.7917,14.3062)[bl]{$b$}
\FAProp(7.,12.)(13.,12.)(-0.9165,){/Straight}{1}
\FALabel(10.,16.02)[b]{$c,u$}
\FAProp(13.,12.)(18.,15.)(0.,){/Straight}{1}
\FALabel(15.2082,14.3062)[br]{$s$}
\FAProp(2.,9.)(7.,12.)(0.,){/Straight}{-1}
\FALabel(4.6888,9.8652)[tl]{$\bar{s}$}
\FAProp(7.,12.)(13.,12.)(1.0832,){/Straight}{-1}
\FALabel(10.,7.9303)[t]{$\bar{c}, \bar{u}$}
\FAProp(13.,12.)(18.,9.)(0.,){/Straight}{-1}
\FALabel(15.3111,9.8652)[tr]{$\bar{b}$}
\FALabel(3.,12.)[r]{$, \, \, \, \, \Gamma_{12} = $}
\FAVert(7.,12.){2}
\FAVert(13.,12.){2}
\end{feynartspicture}
\end{center}
\vspace{-1.5cm}
The vertices in the diagrams for $\Gamma$ and $\Gamma_{12}$
are effective four-quark operators with $\Delta B = 1$,
while the vertex in the diagram for $M_{12}$
is an effective four-quark operator with $\Delta B = 2$.
For $M_{12}$ we have now already the final local operator, whose matrix 
element has to be determined with some non-perturbative QCD-method.
\\
As a next step we rewrite the expression for $\Gamma$ in a form 
that is almost identical to the one of $\Gamma_{12}$.
With the help of the optical theorem $\Gamma$ can be
rewritten (diagramatically: a mirror reflection on the right end of the
decay diagram followed by all possible Wick contractions of the quark lines) 
in
\begin{center}
\begin{feynartspicture}(300,150)(2,1)
\FADiagram{$\Gamma_0$}
\FAProp(2.,15.)(7.,12.)(0.,){/Straight}{1}
\FALabel(4.7916,14.3062)[bl]{$b$}
\FAProp(7.,12.)(13.,12.)(-0.9165,){/Straight}{1}
\FALabel(10.,16.02)[b]{$c,u$}
\FAProp(13.,12.)(18.,15.)(0.,){/Straight}{1}
\FALabel(15.2082,14.3062)[br]{$b$}
\FAProp(7.,12.)(13.,12.)(1.0832,){/Straight}{1}
\FALabel(10.,10.5303)[t]{$s, d$}
\FALabel(2.,12.)[r]{$\Gamma = $}
\FAProp(7.,12.)(13.,12.)(0.,){/Straight}{-1}
\FALabel(10.,13.8)[t]{$\bar{c}, \bar{u}$}
\FAProp(2.,7.)(18.,7.)(0.,){/Straight}{-1}
\FALabel(10.,6.18)[t]{$\bar{s}$}
\FAVert(7.,12.){2}
\FAVert(13.,12.){2}
\FADiagram{$\Gamma_3$}
\FAProp(2.,15.)(7.,12.)(0.,){/Straight}{1}
\FALabel(4.7917,14.3062)[bl]{$b$}
\FAProp(7.,12.)(13.,12.)(-0.9165,){/Straight}{1}
\FALabel(10.,16.02)[b]{$c,u$}
\FAProp(13.,12.)(18.,15.)(0.,){/Straight}{1}
\FALabel(15.2082,14.3062)[br]{$b$}
\FAProp(2.,9.)(7.,12.)(0.,){/Straight}{-1}
\FALabel(4.6888,9.8652)[tl]{$\bar{s}$}
\FAProp(7.,12.)(13.,12.)(1.0832,){/Straight}{-1}
\FALabel(10.,7.9303)[t]{$\bar{c}, \bar{u}$}
\FAProp(13.,12.)(18.,9.)(0.,){/Straight}{-1}
\FALabel(15.3111,9.8652)[tr]{$\bar{s}$}
\FALabel(1.,12.)[r]{$ + \, ... \, + $}
\FALabel(22.,12.)[r]{$ + \, ... $}
\FAVert(7.,12.){2}
\FAVert(13.,12.){2}
\end{feynartspicture}
\end{center}
The first term ($=: \Gamma_0$) corresponds to the decay of a free
$b$-quark, see e.g. \cite{LNO97, LNO98, L00durham} and references therein
for some applications. This term gives the same contribution to all
$b$-hadrons. The lifetime differences we are interested in will only appear 
in subleading terms of this expansion like the second diagram 
($ =: \Gamma_3$), which looks very similar to the diagram for $\Gamma_{12}$.
Counting the mass dimensions of the external lines one can write formally
an expansion of the total decay rate in inverse powers of the heavy quark mass 
$m_b$:
\begin{equation}
\Gamma = \Gamma_0 + \frac{\Lambda}{m_b} \Gamma_1
                  + \frac{\Lambda^2}{m_b^2} \Gamma_2 +
\frac{\Lambda^3}{m_b^3} \Gamma_3 +... \, .
\label{Gammaexp}
\end{equation}
However the expressions for $\Gamma_i$ and  $\Gamma_{12}$ are still non-local,
so we perform a second OPE (OPE II) using the fact that the $b$-quark 
mass is heavier than the QCD scale ($m_b \gg \Lambda_{QCD}$). 
The OPE II is called the heavy quark expansion (HQE) 
\cite{HQE1,HQE2,HQE3,HQE4,HQE5,HQE6,HQE7,HQE8}.
The resulting diagrams for $\Gamma_3$ and $\Gamma_{12}$ look like
the final diagram for $M_{12}$:
\vspace{-1cm}
\begin{center}
\begin{feynartspicture}(300,150)(2,1)
\FADiagram{}
\FAProp(2.,15.)(10.,12.)(0.,){/Straight}{1}
\FALabel(6.5266,14.4244)[b]{$b$}
\FAProp(10.,12.)(18.,15.)(0.,){/Straight}{1}
\FALabel(13.4733,14.4244)[b]{$b$}
\FAProp(2.,9.)(10.,12.)(0.,){/Straight}{-1}
\FALabel(6.4564,9.7627)[t]{$\bar{s}$}
\FAProp(10.,12.)(18.,9.)(0.,){/Straight}{-1}
\FALabel(13.5435,9.7627)[t]{$\bar{s}$}
\FALabel(3,12.)[r]{$\Gamma_3 = $}
\FAVert(10.,12.){3}
\FADiagram{}
\FAProp(2.,15.)(10.,12.)(0.,){/Straight}{1}
\FALabel(6.5266,14.4244)[b]{$b$}
\FAProp(10.,12.)(18.,15.)(0.,){/Straight}{1}
\FALabel(13.4733,14.4244)[b]{$s$}
\FAProp(2.,9.)(10.,12.)(0.,){/Straight}{-1}
\FALabel(6.4564,9.7627)[t]{$\bar{s}$}
\FAProp(10.,12.)(18.,9.)(0.,){/Straight}{-1}
\FALabel(13.5435,9.7627)[t]{$\bar{b}$}
\FALabel(3,12.)[r]{$, \, \,   \, \, \, \, \Gamma_{12} = $}
\FAVert(10.,12.){3}
\end{feynartspicture}
\end{center}
\vspace{-2.5cm}
Now we are left with local four-quark operators 
($\Delta B = 0$ for $\tau$ and $\Delta B = 2$ for $\Gamma_{12}$).
The non-perturbative matrix elements of these operators are 
expressed in terms of decay constants $f_B$ and bag parameters $B$.
In the standard model one gets one operator for $M_{12}$, two 
independent operators for $\Gamma_{12}$ - including the operator
that appears in $M_{12}$ - and e.g. four operators for $\tau (B^+) / \tau(B_d)$
\footnote{This statements hold only at order $1/m_b^3$.} - in 
extensions of the standard model typically more operators arise.
\\
At this stage we would like to make some comments:
\begin{itemize}
\item One can show that in the end no corrections of order $1/m_b$ survive 
      in the total decay rate in Eq. (\ref{Gammaexp}).
\item $\Gamma$ and $\Gamma_{12}$ are expected to be almost free from possible
      new physics contributions, since only light internal particles
      contribute, while $M_{12}$ might easily have large contributions
      from new physics effects. 
      Since allowed new physics contributions to $\Gamma$ and $\Gamma_{12}$ 
      are smaller than the QCD uncertainties we neglect them in the following.
\item The OPE II seems to be theoretically less justified than
      the OPE I ($m_W/m_b \approx 17...19 > 4...10 \approx m_b/\Lambda_{QCD}$),
      but the HQE can be tested directly by comparing experiment and theory
      for the lifetimes.
\item In all the diagrams shown in this section perturbative QCD-corrections
      have to be included! These corrections to the Wilson coefficients
      turned out to be quite sizeable.
\end{itemize}
Summarizing one can state:
the HQE represents a systematic expansion, which can in principle be 
tested by the lifetimes - in that sense it is not a model like 
the quark model.
\section{Introduction - Motivation}
Besides testing our understanding of QCD and determining the standard model
parameters the search for new physics effects is a basic motivation for 
the study of the mixing quantities.
Since $M_{12}$ is sensitive to heavy new internal particles one might start 
with $\Delta M_s$, which is proportional to $f_{B_s}^2 B$ (see next section).
Unfortunately $f_{B_s}$ is hardly known.
To visualise our current unsatisfactory knowledge of the precise value
of the decay constant, we have taken some recent numerical values for 
$f_{B_s}$ from the literature \cite{DM07,Arifa07,HPQCD05,DM06,GL05} 
and calculated the corresponding value of the mass difference $\Delta M_s$.
\\
\begin{equation}
\begin{array}{cccc}
f_{B_s}                          & N_F &   \Delta M_s                 & 
\mbox{deviation from experiment}
\\
\\
193 \pm 06 \,\, \mbox{MeV}  \, \, [17] & 0 & 12.5 \pm 1.4 \,\, \mbox{ps}^{-1} &
- 3.9 \, \, \sigma 
\\
205 \pm 32 \,\, \mbox{MeV}  \, \, [18] & 2 & 14.1 \pm 4.6 \,\, \mbox{ps}^{-1} &
- 0.8 \, \, \sigma 
\\
259 \pm 26 \,\, \mbox{MeV}  \, \, [19] & 3 & 22.5 \pm 5.0 \,\, \mbox{ps}^{-1} &
+ 0.9 \, \, \sigma 
\\
297 \pm 14 \,\, \mbox{MeV}  \, \, [20] & 2 & 30.0 \pm 3.9 \,\, \mbox{ps}^{-1} &
+ 3.1 \, \, \sigma 
\\
341 \pm 32 \,\, \mbox{MeV}  \, \, [21] & 2 & 39.0 \pm 8.2 \,\, \mbox{ps}^{-1} &
+ 2.6 \, \, \sigma 
\end{array}
\end{equation}
Depending on your favorite lattice collaboration you can arrive at theory 
predictions that are smaller, are equal or are higher than the 
experimental value for the mass difference $\Delta M_s$.
This unfortunate situation might be called the {\bf decay constant problem}. 
Here clearly more work has to be done to settle this ignorance 
and moreover the error estimates have to be done with much care.
In the analysis in \cite{LN06} we use the conservative estimate
$f_{B_s} = 240 \pm 40$ MeV.
\\
In order to circumvent the decay constant problem one might try to determine
the ratio  $\Delta M_s / \Delta M_d$. Here the ratio
$|V_{ts}^2/V_{td}^2| \cdot f_{B_s}^2 B_{B_s}/(f_{B_d}^2 B_{B_d})$ 
arises. Although the ratio of the non-perturbative parameters is claimed 
to be theoretically better under control, one is still left
with the uncertainty in the CKM elements, which is of the order
of $40 \%$.
\\
In the ratio $\Gamma_{12}/M_{12}$ the decay constant and the bag parameter
$B$ from $M_{12}$ cancel completely,
schematically one gets
\begin{equation}
\frac{\Gamma_{12}}{M_{12}} = a + b \frac{B_X}{B} + 
{\cal O} \left( \frac{1}{m_b^4}\right) \, .
\end{equation}
We get a term that is completely free of any non-perturbative
uncertainties ($a$) and a term that depends on the ratio of two bag parameters
($b$). If $a > b$ and if the remaining uncertainties are under control
then $\Gamma_{12}/M_{12}$ might be an ideal quantity to search for new 
physics.
Moreover the accuracy in the determination of this ratio can be tested 
directly via the lifetimes, which root on the same theoretical footing.
%
\section{State of the art}
%
In this section we summarize the current status in the theoretical
determination of the lifetimes of the $b$-hadrons and the mixing quantities.
\subsection{The mass difference - $M_{12}$}
Calculating the box diagram with internal top quarks one obtains
 \begin{equation}
        M_{12,q} =  \frac{G_F^2}{12 \pi^2} 
          (V_{tq}^* V_{tb})^2 M_W^2 S_0(x_t)
          {B_{B_q} f_{B_q}^2  M_{B_q}} \hat{\eta }_B \, .
        \end{equation}
The Inami-Lim function $S_0 (x_t = \bar{m}_t^2/M_W^2)$ 
\cite{IL} is the result of the box diagram 
without any gluon corrections. The NLO QCD correction is parameterized by 
$\hat{\eta}_B \approx 0.84$ \cite{BJW}.
The non-perturbative matrix element is parameterized by the 
bag parameter $B$ and the decay constant $f_B$.
\subsection{The decay rate difference - $\Gamma_{12}$}
The calculation of $\Gamma_{12}$ is a little bit more involved since
a second OPE has to be performed.
$\Gamma_{12}$ can be expanded as
\begin{equation}
\Gamma_{12} = \frac{\Lambda^3}{m_b^3} 
\left(\Gamma_3^{(0)} + \frac{\alpha_s}{4 \pi} \Gamma_3^{(1)} + ... \right)
+ 
\frac{\Lambda^4}{m_b^4} 
\left(\Gamma_4^{(0)} + ... \right) + ... \, .
\end{equation}
The $1/m_b$-corrections ($\Gamma_4^{(0)}$) were determined in \cite{BBD96}
and they turned out to be quite sizeable.
NLO QCD-corrections were done for the first time in \cite{BBGLN98}, they also
were quite large. At that time no lattice results were available for all 
appearing four-quark operators, so no real numerical prediction could be made.
The first numerical estimate including NLO-QCD corrections and 
non-perturbative determinations of the appearing four-quark operators
was given in \cite{BL00durham}.
Five years later the QCD-corrections were confirmed and also
subleading CKM structures were included \cite{BBLN03, rome03}.
Unfortunately it turned out that $\Delta \Gamma$ is not well-behaved 
\cite{L04korea}. All corrections are unexpectedly large and they go in 
the same direction.
This problem could be solved by introducing a new operator
basis \cite{LN06}. As an illustration of the improvement
we show the expressions for $\Gamma_{12}/M_{12}$ in the old and the 
new basis:
\begin{eqnarray}
\frac{\Delta \Gamma_s}{\Delta M_s}^{\mbox{Old}} & = &
10^{-4} \cdot
\left[ 0.9  + 40.9 \frac{ B_S'}{B}  - 25.0  \frac{B_R}{B} 
\right] \, ,
\\
\frac{\Delta \Gamma_s}{\Delta M_s}^{\mbox{New}} & = &
10^{-4} \cdot
\left[ 46.2  + 10.6 \frac{ B_S''}{B}  - 11.9  \frac{B_R}{B} 
\right] \, .
\end{eqnarray}
Now the term that is completely free of any non-perturbative uncertainties
is numerical dominant. Moreover the $1/m_b$-corrections became smaller and
undesired cancellations are less pronounced. For more details we refer the
reader to \cite{LN06}.
Currently also $1/m_b$-corrections for the subleading CKM structures
in $\Gamma_{12}$ \cite{DHKY01} 
and $1/m_b^2$-corrections for $\Delta \Gamma_s$ \cite{BGP07}
are available - they are relatively small.
\subsection{Lifetimes}
The lifetime ratio of two $b$-hadrons can be written as
\begin{equation}
\frac{\tau_1}{\tau_2} = 1 + \frac{\Lambda^2}{m_b^2} \Gamma_2 +
\frac{\Lambda^3}{m_b^3} 
\left(\Gamma_3^{(0)} + \frac{\alpha_s}{4 \pi} \Gamma_3^{(1)} + ... \right)
+ 
\frac{\Lambda^4}{m_b^4} 
\left(\Gamma_4^{(0)} + ... \right) + ... \, .
\end{equation}
$\Gamma_2$ vanishes e.g. in $\tau_{B^+}/\tau_{B_d}$,  $\tau_{B_s}/\tau_{B_d}$
and  $\tau_{\Xi_b^+}/\tau_{\Xi_b^0}$ but it survives in  
$\tau_{\Lambda_b}/\tau_{B_d}$.
The sizeable NLO QCD-corrections to the lifetime ratios ($\Gamma_3^{(1)}$)
were determined in \cite{BBGLN02,rome02};
$1/m_b$-corrections  ($\Gamma_4^{(0)}$) and $1/m_b^2$-corrections 
($\Gamma_5^{(0)}$) were calculated in \cite{Petrov04} - 
they are negligible for 
 $\tau_{B^+}/\tau_{B_d}$ and $\tau_{B_s}/\tau_{B_d}$, but they might be 
sizeable for $\tau_{\Lambda_b}/\tau_{B_d}$.
%
\section{Numerical Results}
%
\subsection{Lifetimes}
The theoretically best investigated lifetime ratio is $\tau_{B^+}/\tau_{B_d}$.
One obtains \cite{BBGLN02,rome02}
\begin{equation}
\frac{\tau (B^+)}{\tau (B_d)} = 1.063 \pm 0.027 \, .
\end{equation}
NLO-QCD corrections turned out to be important, while subleading 
$1/m_b$-corrections are negligible.
Some care has to be taken with the arising 
matrix elements of the four-quark operators: it
turned out that the Wilson coefficients of the 
color-suppressed operators are numerically enhanced, 
see \cite{BBGLN02}. But the matrix elements of these operators
are only knwon with large relative errors.
Currently two determinations on the lattice are available 
\cite{Latticelifetime98, Latticelifetime01}. 
\\
For  $\tau_{B_s}/\tau_{B_d}$ large cancellations occur so the ratio is 
expected to be very close to one \cite{BBD96,rome02}
\begin{equation}
\frac{\tau (B_s)}{\tau (B_d)} = 1.00 \pm 0.01 \, .
\end{equation}
Predictions for the $\Lambda_b$ have to be taken with more care.
In that case the NLO-QCD corrections are not complete and only 
preliminary lattice values \cite{Latticelifetime99} are available. 
A typical value quoted in the literature \cite{Tarantino07} is
\begin{equation}
\frac{\tau (\Lambda_b)}{\tau (B_d)} = 0.88 \pm 0.05 \,.
\end{equation}
The lifetime of the doubly heavy meson $B_c$ has been investigated e.g. in
\cite{taubctheory}, but only in LO QCD.
\begin{displaymath}
\tau(B_c)_{\rm LO} = 0.52_{-0.12}^{+0.18} \,  \mbox{ps} \, .
\end{displaymath}
In addition to the b-quark now also the c-quark can decay, giving rise
to the biggest contribution to the total decay rate.
\\
An interesting quantity is the lifetime ratio of the $\Xi_b$-baryons, which 
was investigated in NLO-QCD in \cite{BBGLN02}. This quantity can in principle 
be determined as precise as $\tau_{B^+}/\tau_{B_d}$ ($\pm 3 \%$). 
However, up to now the matrix elements for the $\Xi_b$ baryons are not 
available.
Assuming that the matrix elements for $\Xi_b$ are equal to the ones of
$\Lambda_b$ we can give a rough estimate for the expected lifetime ratio.
In order to get rid of unwanted $s \to u$-transitions we define (following
\cite{BBGLN02})
\begin{equation}
\frac{1}{\bar{\tau} (\Xi_b)} = \bar{\Gamma}  (\Xi_b) 
= \Gamma  (\Xi_b) -  \Gamma  (\Xi_b \to \Lambda_b + X) \, .
\end{equation}
Using the preliminary lattice values \cite{Latticelifetime99} for the 
matrix elements of $\Lambda_b$ we obtain
\begin{equation}
\frac{\bar{\tau} (\Xi_b^0)}{\bar{\tau} (\Xi_b^+)} = 1 - 0.12 \pm 0.02 \pm ???
\, ,
\end{equation}
where $???$ stands for some unknown systematic errors.
As a further approximation we equate $\bar{\tau} (\Xi_b^0)$ 
to $\tau (\Lambda_b)$ - here similar cancellations arise as in 
 $\tau_{B_s}/\tau_{B_d}$ - , so we arrive at the following prediction
\begin{equation}
\frac{\tau (\Lambda_b)}{\bar{\tau} (\Xi_b^+)} = 0.88 \pm 0.02 \pm ??? \, .
\end{equation}
\subsection{Mixing}
The mixing quantities have been investigated in detail in \cite{LN06}, 
numerically we obtain
\begin{eqnarray}
\Delta M_d = 0.53 \pm 0.18 \, \mbox{ps}^{-1} \, ,
&& 
\Delta M_s = 19.3 \pm 6.7 \, \mbox{ps}^{-1} \, ,
\\
\Delta \Gamma_d = (2.67^{+0.58}_{-0.65}) \cdot 10^{-3} \, \mbox{ps}^{-1}  \, ,
&& 
\Delta \Gamma_s = 0.096 \pm 0.039 \, \mbox{ps}^{-1} \, ,
\\
\Delta \Gamma_d / \Gamma_d   = (4.09^{+0.89}_{-0.99}) \cdot 10^{-3} \, ,
&& 
\Delta \Gamma_s / \Gamma_s   = 0.147 \pm 0.060  \, ,
\\
\Delta \Gamma_d / \Delta M_d = (52.6^{+11.5}_{-12.8} ) \cdot 10^{-4} \, ,
&& 
\Delta \Gamma_s / \Delta M_s = (49.7 \pm 9.4) \cdot 10^{-4} \, ,
\\
\phi_d = -0.091^{+0.026}_{-0.038} \, ,
&& 
\phi_s = (4.2 \pm 1.4) \cdot 10^{-3}  \, ,
\\
a_{fs}^d = (-4.8^{+1.0}_{-1.2}) \cdot 10^{-4} \, ,
&& 
a_{fs}^s = (2.06 \pm 0.57) \cdot 10^{-5} \, .
\end{eqnarray}
The predictions for $\Delta \Gamma_d$ and $\Delta \Gamma_d / \Gamma_d$
are obtained \cite{LN06} under the assumption that there are no new physics
contributions in $\Delta M_d$. 
From this list one sees the strong suppression of $\phi$ and $a_{sl}$ in the
standard model.
\section{Experimental status}
\subsection{Lifetimes}
The Heavy Flavor Averaging Group quotes \cite{HFAG07}
the following numbers
\begin{eqnarray}
\frac{\tau (B^+)}{\tau (B_d)} = 1.071 \pm 0.009 \, ,
&& 
\frac{\tau (B_s)}{\tau (B_d)} = 0.939 \pm 0.021 \, ,
\\
\frac{\tau (\Lambda_b)}{\tau (B_d)} = 0.921 \pm 0.036 \, ,
&& 
\tau (B_c) =  0.463 \pm  0.071 \, \mbox{ps} \, .
\end{eqnarray}
From the ratio $\tau_{B^+}/\tau_{B_d}$ it can be seen that the HQE works 
very well. 
$\tau_{B_s}/\tau_{B_d}$ is about 2.9 $\sigma$ below 1, here more precise 
numbers are needed, to see whether there might be some interesting effects.
The situation for the $\Lambda_b$-baryon is not settled yet. First several
theoretical improvements have to be included, second there are two different
experimental numbers on the market \cite{LambdabCDF06, LambdabD007}.
For $B_c$ the number lies in the right ball park, but here also a full 
NLO-QCD calculation would be desireable to make the comparison more 
quantitaive.
Finally we are waiting for a first result for the lifetimes of the 
$\Xi_b$-baryons.
\subsection{Mixing}
The mass differences have been measured with great precision
at LEP, TeVatron and the B factories
\cite{DMS106,DMS206,DMS306,HFAG07}
\begin{eqnarray}
\Delta M_d & = & 0.507 \pm 0.005 \, \mbox{ps}^{-1} \, ,
\\
\Delta M_s & = & 17.77 \pm 0.10 \pm 0.07 \, \mbox{ps}^{-1}\, .
\end{eqnarray}
Due to the uncertainties in the decay constants, theory will not be able 
to achieve a similiar accuracy in the foreseeable future.
\\
For the remaining mixing quantities $a_{sl}$, $\Delta \Gamma$ and $\phi$ we 
do not have measurements yet, but very interesting bounds:
\\
In \cite{dimuonD006,dimuonCDF06} the dimuon-asymmetry was determined
\begin{eqnarray}
a_{sl} & = & 0.582 \, a_{sl}^d + 0.418 \, a_{sl}^s \, ,
\\
a_{sl}^{\mbox{D0}}  & = & (-5.3 \pm 2.5 \pm 1.8) \cdot 10^{-3} \, ,
\\
a_{sl}^{\mbox{CDF}} & = & (+8.0 \pm 9.0 \pm 6.8) \cdot 10^{-3} \, .
\end{eqnarray}
The semileptonic CP asymmetry was also measured directly in
\cite{semiCP07}.
\begin{eqnarray}
a_{sl}^s & = & (1.23 \pm 0.97 \pm 0.17) \cdot 10^{-2} \, .
\end{eqnarray}
Here more precise numbers are needed, because a clear deviation from the 
small standard model value would be be an unambiguous sign for new physics!
The same argument holds for the phase $\phi_s$, while a clean measurement of
$\Delta \Gamma$ is probably best exploited by comparing experiment and theory
for $\Delta \Gamma / \Delta M$.
\\
$\Delta \Gamma$ and $\phi_s$ have been 
determined from an angular analysis in the 
decay $B_s \to J /\psi \phi$:
In the untagged analysis from D0 \cite{untaggedD007}
the following values were obtained
\begin{eqnarray}
\phi_s & = & 0.79 \pm 0.56^{+0.14}_{-0.01} \, ,
\\
\Delta \Gamma & = & 0.17 \pm 0.09 \pm 0.02 \, \mbox{ps}^{-1}\, .
\end{eqnarray}
One has to keep in mind the 4-fold ambiguity in $\phi_s$: with $\phi_s$ also
$-\phi_s$ and $\pi \pm \phi_s$ are solutions!
CDF obtained from the untagged analysis \cite{untaggedCDF07}
\begin{eqnarray}
\Delta \Gamma & = & 0.076^{+0.059}_{-0.063} \pm 0.006 \, \mbox{ps}^{-1}\, .
\end{eqnarray} 
and no bound on $\phi_s$.
\\
CDF also performed a tagged analysis \cite{taggedCDF07}
and obtains confidence regions in the
$\phi_s - \Delta \Gamma$-plane, which differ about 1.5 $\sigma$
from the SM prediction. If they fix $|\Gamma_{12}|$ to the SM
value obtained in \cite{LN06} they get
\begin{eqnarray}
- \phi_s & \in &  [0.24, 1.36]  \cup [1.78,2.90]\, .
\end{eqnarray} 

%
%
%
%
%
%
%
%
\section{New physics models}
\begin{figure}
\includegraphics[width=0.95\textwidth,angle=0]{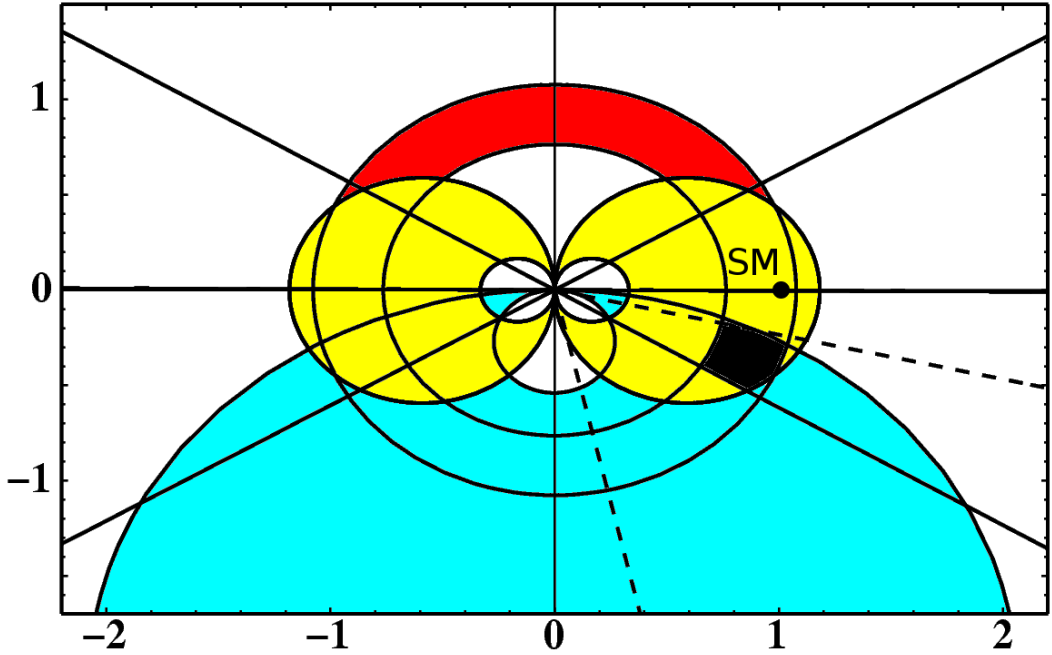}
\caption{Experimental bounds in the complex $\Delta_s$-plane 
  (state: end of 2006).
  The bound from $\Delta M_s$ is given by the red (dark-grey) ring around
  the origin. The bound from $\Delta \Gamma_s / \Delta M_s$ is given
  by the yellow (light-grey) region and the bound from $a_{fs}^s$ is given
  by the light-blue (grey) region. The angle $\phi_s^\Delta$ can be extracted
  from $\Delta \Gamma_s$ (solid lines) with a four fold ambiguity - one bound
  coincides with the x-axis! - or from the angular analysis in
  $B_s \to J / \psi \phi$ (dashed line). If the standard model is valid
  all bounds should coincide in the point (1,0). The current experimental
  situation shows a small deviation, which might become significant, if the
  experimental uncertainties in $\Delta \Gamma_s$, $a_{sl}^s$ and $\phi_s$
  will go down in near future.}\label{boundbandreal}
\end{figure}

%
In the literature many new physics models are applied to the mixing sectors, 
e.g. \cite{KKL07,L07} and references in \cite{LN06}.
%
In \cite{LN06} we have presented a model independent way to determine
new physics effects in the mixing sector.
We assume that new physics does not alter $\Gamma_{12}$ - at least 
not more than the intrinsic QCD uncertainities, but it might have a
considerable effect on $M_{12}$. Therefore we write
\begin{equation}
\Gamma_{12} = \Gamma_{12}^{\mbox{SM}} 
\hspace{1cm}
M_{12} = M_{12}^{\mbox{SM}} \cdot \Delta
\end{equation}
By comparing experiment and theory for the different mixing observables
we get bounds in the complex $\Delta$-plane, see \cite{LN06}.
Taking the solution for $\phi_s$ from the untagged D0 analysis 
in the 4th quadrant - which corresponds 
to a certain choice of the strong phases in the decay
$B_s \to J /\psi \phi$ - and the data that were available at the end of 2006,
we obtained in \cite{LN06} a 2 $\sigma$ deviation from the standard model,
see Fig. (\ref{boundbandreal}). A new analysis is currently in progress.

\underline{\bf Note added:} (taken from \cite{L07elba})
There is sometimes a confusion between the mixing phases $\beta_s$ 
and $\phi_s$, which we would like to adress here. Both numbers are 
expected to be small in the standard model -
$\phi_s =  (0.24 \pm 0.04)^\circ$ and 
$ 2 \beta_s =  (2.2 \pm 0.6)^\circ (= (0.04 \pm 0.01) \mbox{rad} )$, 
but in view of the high future experimental precisions - in particular at
LHCb \cite{CERN08}- a clear distinction 
might be useful.
\\
$ 2 \beta_s := - \mbox{arg} [(V_{tb}V_{ts}^*)^2 /(V_{cb}V_{cs}^*)^2 ]$
is the phase which appears in $b \to c \bar{c} s$ decays of neutral
B-mesons taking possible mixing into account, 
so e.g. in the case $B_s \to J/\psi + \phi$.
$(V_{tb}V_{ts}^*)^2$ comes from the mixing (due to $M_{12}$) and 
$(V_{cb}V_{cs}^*)^2$ comes from the ratio of $b \to c \bar{c} s$ decay 
and $\bar{b} \to \bar{c} c \bar{s}$ amplitudes.
Sometimes $ \beta_s $ is approximated as 
$ 2 \beta_s \approx  - \mbox{arg} [(V_{tb}V_{ts}^*)^2] \approx 
- \mbox{arg} [(V_{ts}^*)^2] $ - the error due to this approximation is on the
per mille level.
\\
$\phi_s := \mbox{arg} [M_{12}/\Gamma_{12}]$ is the phase that appears e.g. in
$a_{fs}^s$. In $M_{12}$ we have again $(V_{tb}V_{ts}^*)^2$, while we have a 
linear combination of  $(V_{cb}V_{cs}^*)^2$, $V_{cb}V_{cs}^*V_{ub}V_{us}^*$ 
and $(V_{ub}V_{us}^*)^2$ in $\Gamma_{12}$. Neglecting the latter two 
contributions - which is not justified - would yield the phase $2 \beta_s$.
\\
New physics alters the phase $-2 \beta_s$ to $\phi_s^\Delta-2 \beta_s$
and the phase $\phi_s$ to $\phi_s^\Delta +  \phi_s$.
If the new physics contribution is sizeable, then in both cases only
$\phi_s^\Delta$ survives, since the standard model phases are very small.
\\
In the tagged analysis CDF \cite{taggedCDF07} introduces the phase
$2 \beta_s$ for which the following relation to the notion in \cite{LN06}
holds $- 2 \beta_s := \phi_s^{\Delta} - 2 \beta_s^{SM} \, .$
%
%
%
%
\section{Outlook}
In this talk we have summarized the current theoretical status of the
lifetimes
of $b$-hadrons and the mixing quantities.
Our main strategy for finding new physics in these quantities is the following:
New physics is expected to have the biggest effects in $M_{12}$, but due 
to the decay constant problem the quantity that comes first in mind - 
$\Delta M$ - seems to be not the best choice. We have argued that
$\Gamma_{12} / M_{12}$ is theoretically very well under control. Therefore
our first choice are the quantities $\Delta \Gamma / \Delta M$, $a_{sl}$ and
$\phi$.
Moreover the theoretical precision in the determination of $\Gamma_{12}$ 
can be tested directly by investigating the lifetimes of $b$-hadrons, because
both quantities rely on the same theoretical footing.

We conclude with a subjective wish-list for theory and experiment:
\begin{itemize}
\item \underline{Perturbative calulations:}
   \begin{itemize}
   \item NLO-QCD corrections  for $\tau(B_c)$
   \item complete NLO-QCD corrections for $\tau(\Lambda_b)$
   \item $\Gamma_4^{(1)}$ for $\Gamma_{12}$
   \item $\Gamma_3^{(2)}$ for $\Gamma_{12}$
   \end{itemize}
\item \underline{Non perturbative calculations:}
   \begin{itemize}
   \item matrix elements for $\tau(B^+)/ \tau(B_d)$
   \item matrix elements for $\tau(\Lambda_b)$ and $\tau(\Xi_b)$
   \item precise and relieable values for the decay constants
   \item $1/m$-operators for $\Gamma_{12}$, a first step in that direction
         has been performed in \cite{Siegen07}
   \end{itemize}
\item \underline{Experiment: (ranked)}
   \begin{itemize}
   \item[1)] Precise values for $a_{sl}$ and $\phi_s$
   \item[2)] Precise values for $\tau(B_s)/ \tau(B_d)$ and $\tau(\Lambda_b)$ 
             and   $\Delta \Gamma$ --- a first value for $\tau(\Xi_b)$
   \end{itemize}
\end{itemize}

\begin{theacknowledgments}
I would like to thank Martin Beneke, Gerhard Buchalla, Christoph Greub,
Andreas Jantsch, Naoko Kifune, Jisuke Kubo, Heiko Lacker and Uli Nierste for
the pleasant collaboration on topics covered in this talk 
and the organizers of this workshop for their perfect work.

\end{theacknowledgments}

\section{Appendix}
\doingARLO[\bibliographystyle{aipproc}]
          {\ifthenelse{\equal{\AIPcitestyleselect}{num}}
             {\bibliographystyle{arlonum}}
             {\bibliographystyle{arlobib}}
          }

\bibliography{Lenz}

\end{document}